\documentclass[epj]{svjour}

\usepackage{graphicx,epsfig}
\usepackage{amsmath,amssymb,amsfonts,bm,color}

\RequirePackage[numbers,sort&compress]{natbib}
\journalname{Eur. Phys. J. C}

\newcommand{\be}{\begin{equation}}
\newcommand{\ee}{\end{equation}}
\newcommand{\bea}{\begin{eqnarray}}
\newcommand{\eea}{\end{eqnarray}}
\newcommand{\beas}{\begin{eqnarray*}}
\newcommand{\eeas}{\end{eqnarray*}}
\newcommand{\hhh}{$\vphantom{\int_0^1}$}

\newcommand{\alphacc}{\alpha_S^{\bar{c}c}}
\newcommand{\alphabb}{\alpha_S^{\bar{b}b}}

\newcommand{\Ccc}{C_0^{\bar{c}c}}
\newcommand{\Cbb}{C_0^{\bar{b}b}}

\newcommand{\veS}{{\vec S}}
\renewcommand{\vec}{\bm}
\newcommand{\vep}{\vec{p}}
\newcommand{\veq}{\vec{q}}

\graphicspath{{Figs.dir/}}

\begin{document}

\title{$X(6200)$ as a compact tetraquark in the QCD string model}

\author{A.V. Nefediev\inst{1,2,}\thanks{E-mail: nefedev.av@mipt.ru}}
\institute{
Moscow Institute of Physics and Technology, 141700, Institutsky lane 9, Dolgoprudny, Moscow Region, Russia
\and
P.N. Lebedev Physical Institute of the Russian Academy of Sciences, 119991, Leninskiy Prospect 53, Moscow, Russia
}

\date{}

\abstract{
Recently the LHCb Collaboration announced the first observation of nontrivial structures in the double-$J/\psi$ mass spectrum in the mass range 6.2-7.2 GeV, and a theoretical coupled-channel analysis of these data performed in Phys. Rev. Lett. {\bf 126}, 132001 (2021) evidenced the existence of a new state $X(6200)$ close to the double-$J/\psi$ threshold. Although its molecular interpretation seems the most plausible assumption, the present data do not exclude an admixture of a compact component in its wave function, for which a fully-charmed compact tetraquark is the most natural candidate. It is argued in this work that the QCD string model is compatible with the existence of a compact $cc\bar{c}\bar{c}$ state bound by QCD forces just below the double-$J/\psi$ threshold. A nontrivial interplay of the quark dynamics associated with this compact state and the molecular dynamics provided by soft gluon exchanges between $J/\psi$ mesons is discussed and the physical $X(6200)$ is argued to be a shallow bound state, in agreement with
the results of the aforementioned coupled-channel analysis of the LHCb data. 
}



\maketitle

\section{Introduction}

In 2020 the LHCb Collaboration announced the first measurement of the double-$J/\psi$ production in the proton-proton collisions in a rather wide energy range from the double-$J/\psi$ threshold at 6.2 GeV and up to approximately 9 GeV \cite{Aaij:2020fnh}. The measured line shape demonstrates a striking behaviour which was reported by LHCb to be compatible with two peaking structures: a narrow peak at approximately 6.9 GeV and a broad hump just above the production threshold. This result immediately attracted a lot of attention of the community and many explanations were suggested, mainly for the narrow peak which was argued to be the manifestation of a fully-charmed tetraquark state located nearby \cite{Bedolla:2019zwg,Deng:2020iqw,Wang:2020ols,Yang:2020rih,Jin:2020jfc,Lu:2020cns,Becchi:2020uvq,Wang:2020gmd,Albuquerque:2020hio,Sonnenschein:2020nwn,Liu:2020lpw,Giron:2020wpx,Richard:2020hdw,Chen:2020aos,Lu:2020qmp,Barabanov:2020jvn,Chao:2020dml,Eichmann:2020oqt,Yang:2020atz,Maciula:2020wri,Wang:2020dlo,Feng:2020riv,Zhao:2020nwy,Gordillo:2020sgc,Faustov:2020qfm,Weng:2020jao,Zhang:2020xtb,Zhu:2020xni,Guo:2020pvt,Feng:2020qee,Cao:2020gul,Rossi:2020ezg,Jin:2020yjn,Wan:2020fsk,Yang:2020wkh,Huang:2020dci,Zhao:2020zjh,Goncalves:2021ytq,Lucha:2021mwx,Albuquerque:2021erv,Faustov:2021hjs,Ke:2021iyh,Huang:2021vtb,Yang:2021hrb,Mutuk:2021hmi,Li:2021ygk,Richard:2021nvn}. Meanwhile, many double-charmonium thresholds reside in this energy region that implies that a naive analysis based on the Breit-Wigner distributions has to be disregarded in favour of an appropriate coupled-channel approach. In particular, it was argued in Ref.~\cite{Dong:2020nwy} that the measured line shape can be naturally explained in terms of minimalistic coupled-channel models with the minimal number of channels composed of vector charmonia. It has been found then that, while the position of the pole(s) responsible for the narrow structure in the signal is only vaguely fixed by the existing data, there exists a prediction robust against variations of the coupled-channel model employed and as such regarded as reliable. Namely, all models used with the parameters directly fitted to the data demonstrated the existence of a pole close to the physical region in the energy complex plane near the double-$J/\psi$ threshold. This finding was independently confirmed later in Ref.~\cite{Liang:2021fzr}. 
The corresponding physical state was named $X(6200)$ and its possible quantum numbers were fixed to be $0^{++}$ or $2^{++}$. The exact position of its pole could not be strictly localised from the present data, so the $X(6200)$ can be a bound, virtual or above-threshold resonance state. Nevertheless, as stated above, its position very close to the double-$J/\psi$ threshold was confirmed in different models and by an independent calculation. Thus a natural question arises about the nature of this state. The most plausible explanation compatible with the data would be its molecular interpretation argued in Ref.~\cite{Dong:2020nwy}. Meanwhile, one of the models employed to predict this state does not exclude an admixture of a compact component in its wave function, for the latter a compact fully-charmed tetraquark being the most natural candidate (the scalar tetraquark will be the central object of this study since the tensor one lies noticeably higher). Such a situation is not unique in the physics of exotic states with heavy quarks since, for example, the patriarch $X(3872)$ requires such a compact component which allows one to explain its observed properties. The most natural candidate for the latter is the $\chi_{c1}'$ generic quarkonium. Although quark models predict its mass to lie about 100 MeV above the relevant neutral $D\bar{D}^*$ threshold, a strong interaction with this channel brings the pole incredibly close to the threshold.
Similar mechanisms may be operative in the $X(6200)$, too. However, in this case, the coupling to continuum channels implies not a relatively ``simple'' string breaking mechanism through a creation of the light quark-antiquark pair, but more involved rearrangements among four heavy quarks. Therefore, it would be natural to expect the bare mass of the corresponding compact fully-charmed tetraquark predicted by the quark model to lie relatively close to the observed $X(6200)$ mass, that is to the double-$J/\psi$ threshold.

The literature on the fully-charmed tetraquarks is quite rich, and the existing predictions for the lowest scalar state form three main subgroups. The models and approaches which enter the first subgroup predict a light $cc\bar{c}\bar{c}$ tet\-ra\-qu\-ark with the mass lying in the range, roughly, 5.9-6.0 GeV \cite{Bedolla:2019zwg,Wang:2020ols,Berezhnoy:2011xn,Wu:2016vtq,Wang:2017jtz,Debastiani:2017msn,Chen:2020lgj}. Even a lighter ground-state $0^{++}$ tetraquark with the mass $5.3\pm 0.5$~GeV is predicted in Ref.~\cite{Heupel:2012ua}. The second subgroup, which is basically as populated as the first one, predicts a relatively heavy tetraquark at approximately 6.4-6.5 GeV \cite{Deng:2020iqw,Jin:2020jfc,Albuquerque:2020hio,Wang:2020dlo,Zhao:2020nwy,Gordillo:2020sgc,Zhang:2020xtb,Albuquerque:2021erv,Mutuk:2021hmi,Chen:2016jxd,Liu:2019zuc}. These calculations received an additional attention of the community after the LHCb result on the double-$J/\psi$ spectrum was announced since they were supposed to explain the enhancement at 6.9 GeV claimed by LHCb. However, neither the above light nor heavy tetraquark can be employed to explain the compact component of the $X(6200)$. In the meantime, there exist predictions for the ground-state tetraquark with the mass around 6.2 GeV. They form a third, the least populated, subgroup. In what follows we refer to two such predictions made using utterly different approaches. Namely, the mass of the lightest scalar fully-charmed tetraquark is predicted to be approximately 6.19 GeV (see Table~\ref{tab:4Q} below for a more accurate value) in the approach of Ref.~\cite{Karliner:2016zzc} based on a phenomenological fit to the known masses of hadrons with heavy quarks. On the contrary, a profound and technically involved relativistic quasipotential quark model is employed in Ref.~\cite{Faustov:2020qfm} to predict exactly the same mass of this tetraquark state (see Table~\ref{tab:4Q}). In addition, we refer to the phenomenological work \cite{Giron:2020wpx} where the mass of the lightest fully-charmed tetraquark was predicted to take a slightly higher value of 6.26 GeV which nevertheless qualitatively falls into the same ballpark of predictions, together with those from Refs.~\cite{Faustov:2020qfm,Karliner:2016zzc}, consistent with the position of the pole of the state $X(6200)$ extracted from the LHCb data in Ref.~\cite{Dong:2020nwy}.

In this work we calculate the mass of the lightest sca\-lar fully-charmed tetraquark in the framework of the QCD string model which is known to be quite successful and useful in studies of various properties of hadrons composed of heavy and light quarks. An attractive feature of this model is that it operates with very appealing and intuitively clear entities such as the string formed by nonperturbative gluons between coloured objects, the colour Coulomb potential between them, various spin-dependent interactions similar to those one is familiar from atomic physics, and so on. For the origins of the model see the review \cite{DiGiacomo:2000irz} on the Field Correlators Method which provides a necessary connection between the phenomenological QCD string model and the fundamental properties of the QCD vacuum. Various aspects of this model applications in hadronic physics are discussed in the lecture notes \cite{Simonov:1999qj}.

The approach to tetraquarks employed in this work is similar in spirit to that adopted in Ref.~\cite{Giron:2020wpx} (including the disclaimer that it is not intended to compete with detailed and comprehensive calculations of the tetraquark spectra), however it brings underlying dynamics into consideration and relates various contributions to the tet\-ra\-qu\-ark mass with such well understood quantities as the string tension responsible for the nonperturbative dynamics and the colour Coulomb potential for the perturbative interaction. The consideration below is based on the diquark-antidiquark picture which is the most popular approach to tetraquarks used in many works, including Refs.~\cite{Faustov:2020qfm,Karliner:2016zzc,Giron:2020wpx} selected here for comparison. Importantly, the calculation made in this work relies entirely on the values of all parameters of the model found previously in Ref.~\cite{Kalashnikova:2016bta} from the studies of the low-lying generic quark-antiquark charmonia and bottomonia. Therefore, the results obtained in this work can be regarded as parameter-free predictions of the QCD string model for the lightest scalar fully-charmed and fully-bottomed tetraquarks. In case of the $cc\bar{c}\bar{c}$ state, the predicted mass agrees surprisingly well with both the results of Refs.~\cite{Faustov:2020qfm,Karliner:2016zzc,Giron:2020wpx} mentioned above and the $X(6200)$ pole position extracted in Ref.~\cite{Dong:2020nwy}. 

If the wave function of the $X(6200)$ is a mixture of a compact component, which can be associated with the compact tetraquark discussed above, and a molecular do\-ub\-le-$J/\psi$ component, a nontrivial interplay of these two dynamics is to take place and affect the properties of the $X(6200)$, especially if the interaction between the $J/\psi$'s can also form a near-threshold pole. The latter question is addressed in detail in a recent work \cite{Dong:2021lkh} and the conclusion has been made that indeed the soft gluon exchanges between the two $J/\psi$ mesons, which hadronise in the form of two-pion and two-kaon exchanges, are capable of binding the two-$J/\psi$ system and producing a near-threshold pole. As a result quite a peculiar situation takes place when several poles generated by different dynamics coexist in the near-threshold region that may result in a highly nontrivial interplay between them \cite{Baru:2010ww,Hanhart:2011jz,Guo:2016bjq}. Here we employ the approach developed in Ref.~\cite{Baru:2010ww} to investigate the behaviour of the poles generated by the quark and molecular dynamics individually and demonstrate how the single-pole scenario for the physical $X(6200)$ advocated in Ref.~\cite{Dong:2020nwy} as a result of the coupled-channel analysis of the LHCb data can be realised. 

The paper is organised as follows. In Sect.~\ref{sec:4c} we employ the QCD string model to evaluate the mass of the lightest fully-charmed tetraquark state and find it just below the double-$J/\psi$ threshold. Then, in Sect.~\ref{sec:interplay}, we investigate an interplay of this bound state pole due to the quark dynamics with the pole generated by the molecular dynamics discussed in Ref.~\cite{Dong:2021lkh}. Finally, we discuss the results obtained in Sect.~\ref{sec:discussion}.

\section{Compact fully-charmed tetraquark in the QCD string model}
\label{sec:4c}

\subsection{Master Shr{\"o}dinger equation and the ground-state solution}

As an important prerequisite consider a nonrelativistic Hamiltonian for two spinless particles of the mass $m$ interacting via a linear potential and a Coulomb force,
\be
H=2m+\frac{p_r^2}{m}+\sigma r-\frac{\alpha}{r},
\label{H}
\ee
where, aiming at the lowest state in the spectrum, we set the angular momentum $l=0$. 

The corresponding Schr{\"o}dinger equation for the radial wave function can be written in the form
\be
\left(-\frac{d^2}{dx^2}+x-\frac{\lambda}{x}\right)\chi^{(\lambda)}(x)=a^{(\lambda)}\chi^{(\lambda)}(x),
\label{Seq}
\ee
where $x$ is a dimensionless coordinate and 
\be
\lambda=\alpha\left(\frac{m}{\sqrt{\sigma}}\right)^{2/3}
\label{lambda}
\ee 
is the reduced strength of the Coulomb potential \cite{Kalashnikova:2000tg,Kalashnikova:2001ig}. The radial wave function $\chi^{(\lambda)}(x)$ is normalised as
\be
\int_0^\infty |\chi^{(\lambda)}(x)|^2 x^2 dx=1.
\ee

In order to solve the Schr{\"o}dinger equation (\ref{Seq}) for the ground state we employ a variational method with the Gaussian test wave function, 
\be
\chi_0^{(\lambda)}(x)=\left(\frac{16\beta^3}{\pi}\right)^{1/4}e^{-\frac12\beta x^2},
\ee
that yields for the ground state energy the expression
\be
a_0^{(\lambda)}(\beta)=\frac32\beta+\frac{2}{\sqrt{\pi\beta}}-2\lambda\sqrt{\frac{\beta}{\pi}},
\ee
which should be minimised with respect to $\beta$,
\be
\frac{\partial a_0^{(\lambda)}(\beta)}{\partial\beta}_{|\beta=\beta_0}=0,\quad a_0^{(\lambda)}=a_0^{(\lambda)}(\beta_0).
\label{difa}
\ee

\begin{figure}[t]
\centerline{\epsfig{file=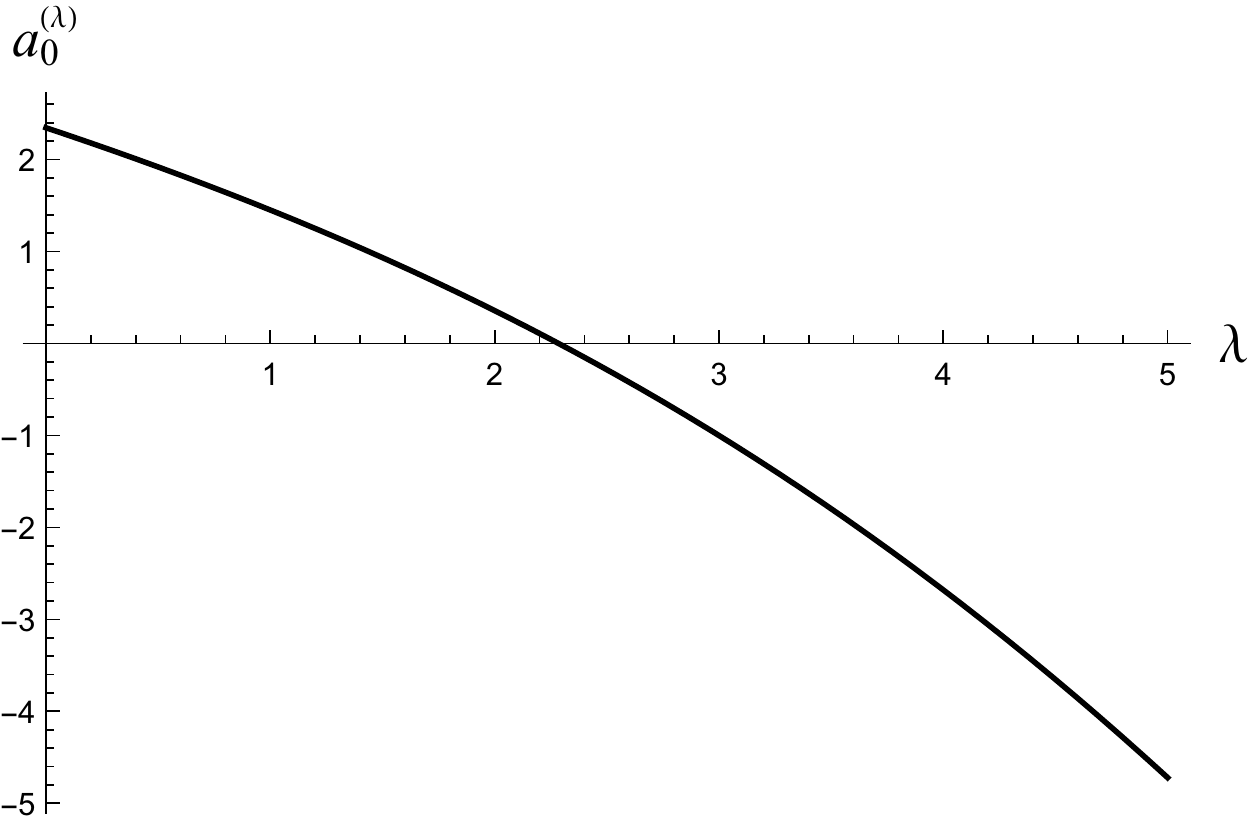, width=0.8\columnwidth}}
\caption{Dependence of the ground state eigenvalue $a_0^{(\lambda)}$ on the reduced Coulomb potential strength $\lambda$.}
\label{fig:alambda}
\end{figure}

Although this procedure can be performed in quadratures in closed form, the result looks bulky and we refrain from quoting it here. Instead, the dependence of $a_0^{(\lambda)}$ on $\lambda$ is shown graphically in Fig.~\ref{fig:alambda}. This procedure provides a very accurate approximation for the exact results that can be verified by setting $\lambda=0$ and comparing the approximate variational value, 
\be
a_0^{(\lambda=0)}\mbox{[var]}=3\left(\frac{3}{2\pi}\right)^{1/3}\approx 2.35,
\ee
with the exact result given by the first zero of the Airy function $Ai$,
\be
Ai(-a_0^{(\lambda=0)}\mbox{[exact]})=0, \quad a_0^{(\lambda=0)}\mbox{[exact]}\approx 2.34.
\ee

The dependence $a_0^{(\lambda)}$ depicted in Fig.~\ref{fig:alambda} implies the existence of two regimes in the Schr{\"o}dinger equation (\ref{Seq}): (i) the regime of small $\lambda$'s when confinement dominates and (ii) the regime of large $\lambda$'s when the Coulomb interaction dominates. The boundary value $\lambda_0$ such that 
$a_0^{(\lambda_0)}=0$ is $\lambda_0\approx 2.28$. It is instructive then to use Eq.~(\ref{lambda}) to estimate the boundary value of the quark mass $m_0$. To this end we set $\sigma=0.16$~GeV and $\alpha= 0.5$ which are close to the phenomenologically adequate values of the fundamental string tension $\sigma_f$ and the strong coupling constant $\alpha_S$ (see also Table~\ref{tab:param} for the actual values of parameters used in calculations below) to find 
\be
m_0\simeq \left(\frac{\lambda_0}{\alpha_S}\right)^{3/2}\sqrt{\sigma}\approx 3.9~\mbox{GeV}.
\ee

Therefore, since the found boundary value exceeds the mass of the charm quark but is smaller than the mass of the bottom quark (see Table~\ref{tab:param} for the values of the parameters adopted in this work), 
\be
m_c<m_0<m_b,
\label{mcm0mb}
\ee
we expect that different regimes, as explained above, are realised in the charmonium and bottomonium diquarks (antidiquarks). 

\begin{figure}[t]
  \centerline{\epsfig{file=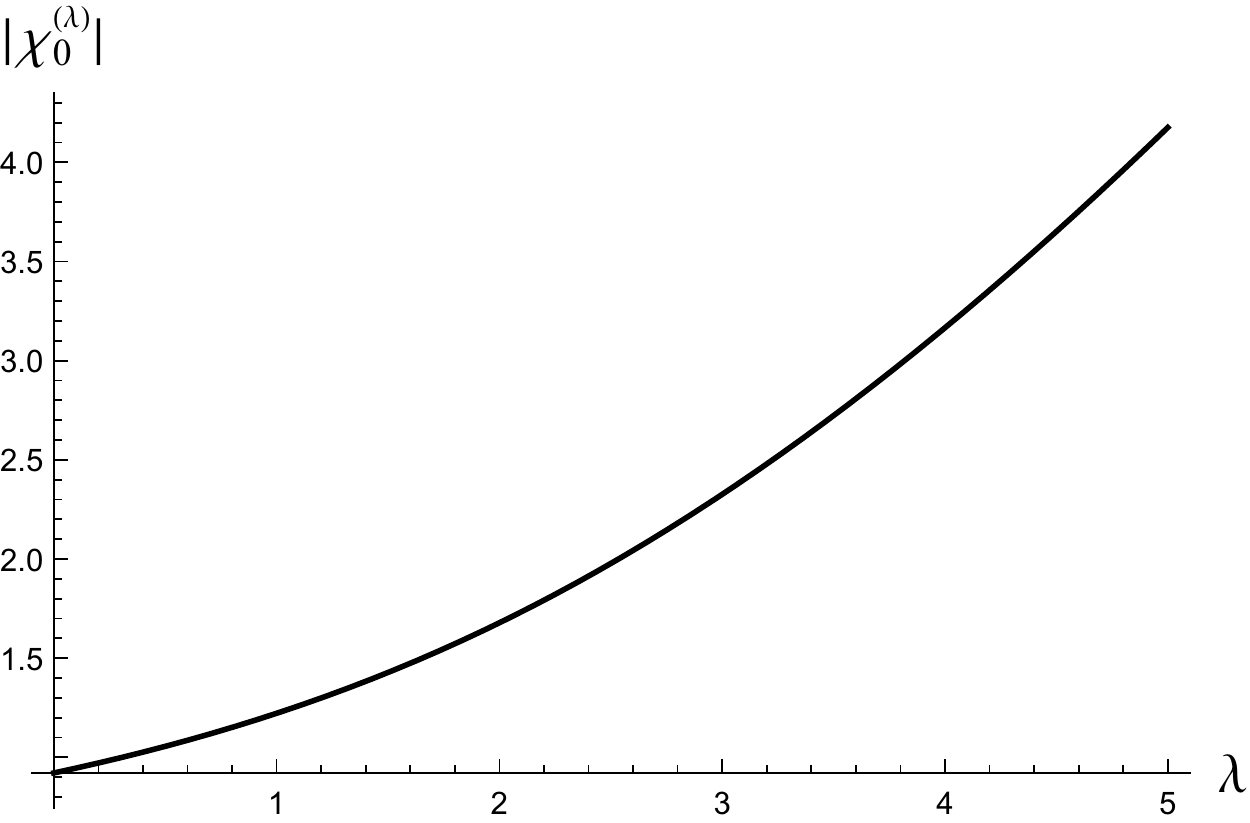, width=0.8\columnwidth}}
  \caption{Dependence of the ground state wave function at the origin $|\chi_0^{(\lambda)}(0)|$ on the reduced Coulomb potential strength $\lambda$.}
    \label{fig:chi0}
    \end{figure}

With the eigenvalue $a_0^{(\lambda)}$ just found, the ground state energy for the Hamiltonian (\ref{H}) takes the form
\be
M_0=2m+a_0^{(\lambda)}\left(\frac{\sigma^2}{m}\right)^{1/3},
\label{Mass0}
\ee
and the value of the radial wave function at the origin is
\be
|R_0(0)|^2=\frac{m\sigma}{4\pi}|\chi_0^{(\lambda)}(0)|^2=m\sigma\left(\frac{\beta_0}{\pi}\right)^{3/2}.
\ee
The analytic formula for the $\lambda$-dependence of $|\chi^{(\lambda)}(0)|$ is also omitted for simplicity --- see the plot of this dependence in Fig.~\ref{fig:chi0}.

\subsection{The string profile and the model parameters}
\label{sec:profile}

\begin{table*}[t]
\begin{center}
\begin{tabular}{|c|c|c|c|c|c|c|c|}
\hline
Parameter&$m_c$, GeV&$m_b$, GeV&$\sigma_f$, GeV$^2$ &$\alphacc$&$\alphabb$&$C_0^{(cc)}$, MeV&$C_0^{(bb)}$, MeV\\
\hline
Value &1.67&4.78&0.16 &0.54&0.42&369&50\\
\hline
\end{tabular}
\end{center}
\caption{Parameters of the model fixed from the fits to the spectrum of ordinary charmonium and bottomonium in Ref.~\cite{Kalashnikova:2016bta}. }\label{tab:param}
\end{table*}

\begin{table*}[t]
\begin{center}
  \begin{tabular}{|c|c|c|c|c|c|c|}
  \hline
  Meson  \hhh         &$\eta_c(1S)$&$J/\psi(1S)$  &$h_c(1P)$&$\chi_{c_1}(1P)$&$\chi_{c_0}(1P)$&$\chi_{c_2}(1P)$\\
  \hline
  $J^P$   \hhh     &$0^{-+}$    &$1^{--}$      &$1^{+-}$ &$1^{++}$        &$0^{++}$        &$2^{++}$\\
  \hline
  $^{2S+1}L_J$ \hhh   &${}^1S_0$&${}^3S_1$&${}^1P_1$&${}^3P_1$&${}^3P_0$&${}^3P_2$\\
  \hline
  Theor., MeV    \hhh  \cite{Kalashnikova:2016bta}   &2981&3104&3528&3514&3449&3552\\
  \hline
  Exp., MeV \hhh \cite{Zyla:2020zbs} &2984&3097&3525&3511&3415&3556\\
  \hline
  \end{tabular}
  \end{center}
  \caption{Masses of the low-lying $S$- and $P$-wave $\bar{c}c$ mesons calculated in Ref.~\cite{Kalashnikova:2016bta}.}\label{tab:cc}
\end{table*}

\begin{table*}
\begin{center}
  \begin{tabular}{|c|c|c|c|c|c|c|}
  \hline
  Meson \hhh&$\eta_b(1S)$&$\Upsilon(1S)$&$h_b(1P)$&$\chi_{b_1}(1P)$&$\chi_{b_0}(1P)$&$\chi_{b_2}(1P)$\\
  \hline
  $J^P$ \hhh&$0^{-+}$&$1^{--}$&$1^{+-}$&$1^{++}$&$0^{++}$&$2^{++}$\\
  \hline
  $^{2S+1}L_J$  &${}^1S_0$\hhh&${}^3S_1$&${}^1P_1$&${}^3P_1$&${}^3P_0$&${}^3P_2$\\
  \hline
  Theor., MeV \hhh\cite{Kalashnikova:2016bta}&9394&9459&9902&9895&9871&9912\\
  \hline
  Exp., MeV \hhh\cite{Zyla:2020zbs} &9398&9460&9899&9893&9859&9912\\
  \hline
  \end{tabular}
  \end{center}
  \caption{Masses of the low-lying $S$- and $P$-wave $\bar{b}b$ mesons  calculated in Ref.~\cite{Kalashnikova:2016bta}.}\label{tab:bb}
  \end{table*}

\begin{figure}[t]
  \centerline{\epsfig{file=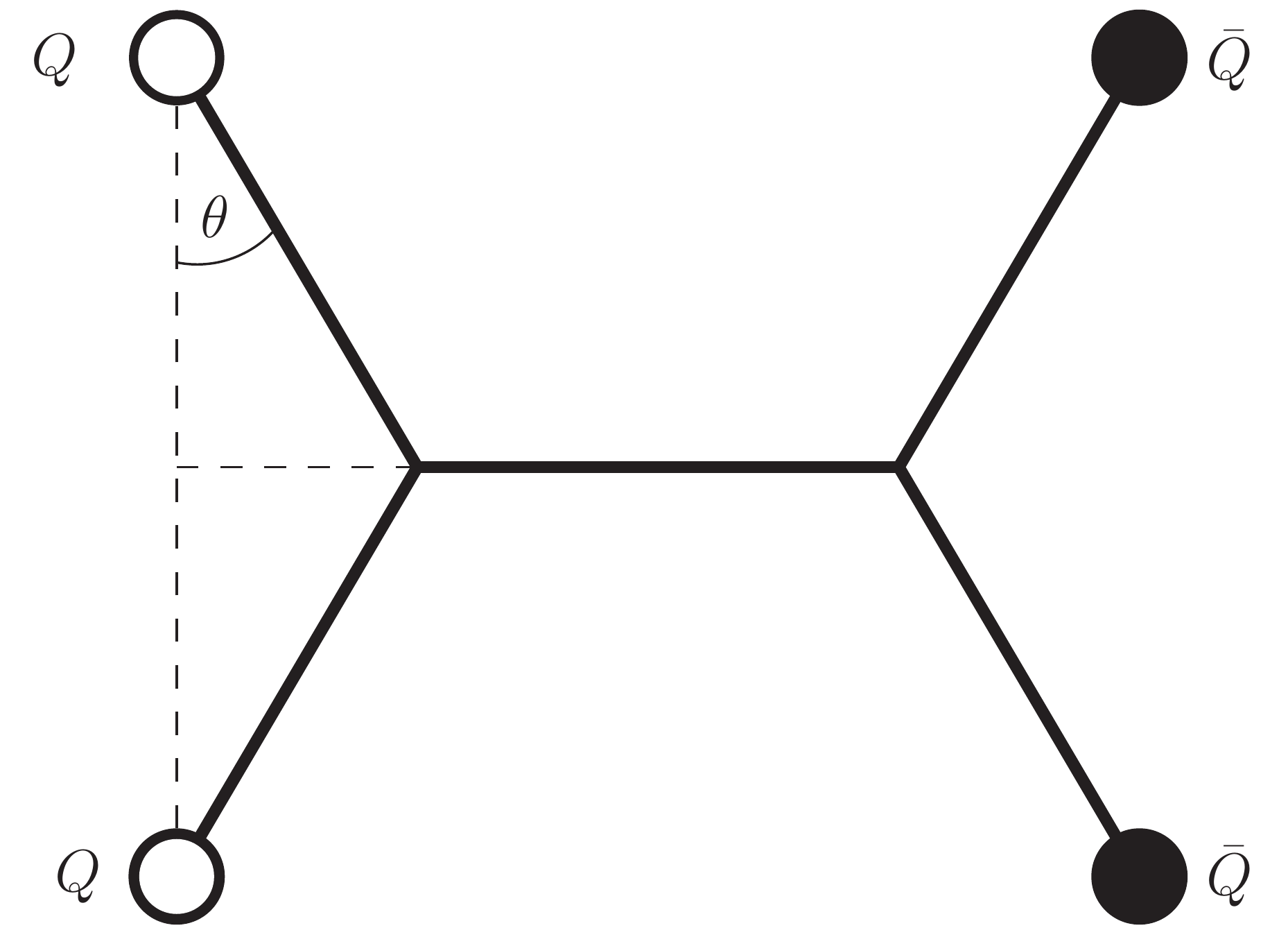, width=0.6\columnwidth}}
  \caption{The string profile for the tetraquark $(QQ)(\bar{Q}\bar{Q})$ made of two heavy quarks and two heavy antiquarks.}
  \label{fig:stringprofile}
  \end{figure}

The string profile for a compact tetraquark in the QCD string model is depicted in Fig.~\ref{fig:stringprofile}. The lowest state in the spectrum corresponds to not excited string degrees of freedom, so that a straight-line ansatz for the string segments connecting coloured objects provides a good approximation for the shape~\cite{Dubin:1994vn,Dubin:1995vw}. The string Y-shaped branching points known as junctions may possess their own dynamics \cite{Kalashnikova:1995kb,Kalashnikova:1996px}, however such modes bring additional energy to the system which appears of the same order as that associated with the string vibrations --- both result in approximately 1 GeV gap between the ground-state energy of the string and its hybrid excitations \cite{Simonov:1999qj}. It is easy to ensure then that the ground-state string configuration corresponds to the minimum of the total string length (this approximation is well justified for heavy quarks) that requires that the three string segments intersect at each junction with the equal angles equal to $2\pi/3$ each. For a three-quark baryon the corresponding point inside the triangle is known as the Torricelli point \cite{Carlson:1982xi}. The tetraquark string configuration possesses two such junctions, so that the tetraquark from Fig.~\ref{fig:stringprofile} is viewed as a diquark--antidiquark system connected by a fundamental string. Such a picture allows one to proceed with the spectrum calculations in a two-step way: the diquark and antidiquark formation and their binding to the tetraquark. The tetraquark with the minimal ground-state energy has all angular momenta between quarks equal to zero. Then, since the colour wave function of the diquark, $\varepsilon_{\alpha\beta\gamma}Q^\beta Q^\gamma$, is antisymmetric and so is the colour wave function of the antidiquark, to comply with the Fermi statistics, their spin wave functions need to be symmetric, so that
$S_{(QQ)}=S_{(\bar{Q}\bar{Q})}=1$ (with $Q$ $(\bar{Q})$ for a heavy quark (antiquark)). 

The parameters of the model are fully fixed from the fit to the spectrum of ordinary charmonium and bottomonium performed in Ref.~\cite{Kalashnikova:2016bta}. They are listed in Table~\ref{tab:param}. To assess the accuracy of the model, in Tables~\ref{tab:cc} and \ref{tab:bb}, we quote the masses of the low-lying $\bar{c}c$ and $\bar{b}b$ states obtained in Ref.~\cite{Kalashnikova:2016bta} in the framework of the QCD string model and with the help of the parameters from Table~\ref{tab:param}. No further fine tuning of the parameters is done in this work, so that the masses of the tetraquarks calculated here come as pure predictions of the QCD string model.

\subsubsection{Diquark (antidiquark)}

\begin{table*}[t]
\begin{center}
\begin{tabular}{|c|c|c|c|c|c|}
\hline
& $\lambda_Q$ & $\beta_Q$ & $a_0^{\lambda_Q}$& $m_{(QQ)}$ (this work), GeV & $m_{(QQ)}$ \cite{Karliner:2016zzc}, GeV \\
\hline
$(cc)/(\bar{c}\bar{c})$ & 0.98 &0.75&1.47&3.31 & 3.2041\\
\hline
$(bb)/(\bar{b}\bar{b})$ & 1.54 &0.95&0.89&9.65 & 9.7189\\
\hline
\end{tabular}
\end{center}
\caption{Values of various auxiliary parameters for the charmonium and bottomonium diquark systems and the effective mass of the corresponding diquark.}\label{tab:QQ}
\end{table*}

The diquark $(QQ)$ (antidiquark $(\bar{Q}\bar{Q})$) can be viewed as a nonrelativistic system bound by the potential
\be
V_{(QQ)}(r)=\sigma' r-C_{\bar{3}}\frac{\alpha_S}{r}+V_{SS}^{(QQ)}(r),
\label{VQQ}
\ee
where $C_{\bar{3}}=2/3$ is the colour Casimir operator in the antitriplet representation of the colour $SU(3)$ group (for convenience, the sign of $C_{\bar{3}}$ is taken into account explicitly in Eq.~(\ref{VQQ})), $\sigma_f$ is the fundamental string tension, and $r$ is the separation between the two quarks $Q$ (antiquarks $\bar{Q}$). The effective string tension $\sigma'=\sigma_f\cos\theta$ (see Fig.~\ref{fig:stringprofile}) provides a smooth interpolation of the confining potential from $V_{\rm conf}(r)=\sigma_f r$ for $\theta=0$, when the two string links between the quarks and the string junction form a straight line connecting the quarks (antiquarks), to $V_{\rm conf}(r)=0$ for $\theta=\pi/2$, when the quarks (antiquarks) sit on top of each other. For the most energetically favourable Y-shaped profile of the string in the diquark (antidiquark) subsystem discussed in the beginning of Sect.~\ref{sec:profile}, one has $\theta=\pi/6$, so that $\sigma'=\sigma_f\cos(\pi/6)=(\sqrt{3}/2)\sigma_f$ and, therefore, the attraction between the quarks due to the confining interaction is somewhat reduced compared to the straight-line string configuration. 

The hyperfine interaction is 
\be
V_{SS}^{(QQ)}(r)=\frac{8\pi C_{\bar{3}}\alpha_S}{9m_Q^2}(\veS_1^{Q}\veS_2^{Q})\delta^{(3)}(r),
\label{VSSQ}
\ee
which has the form of the standard Eichten-Feinberg-Gro\-mes potential \cite{Eichten:1980mw,Gromes:1984ma}, and $m_Q$ is the mass of the heavy quark (antiquark). 
Then, for  $S_{(QQ)}=S_{(\bar{Q}\bar{Q})}=1$ and $L_{(QQ)}=L_{(\bar{Q}\bar{Q})}=0$, it is straightforward to find that
\be
\langle S_{(QQ)}=1|\veS_1^{Q}\veS_2^{Q}|S_{(QQ)}=1\rangle=\frac14
\ee
and the other spin-dependent potentials vanish for the considered quantum numbers of the diquark (antidiquark) system.  

We, therefore, use the general solution of the Sch{\"o}dinger equation (\ref{Seq}) quoted in Eq.~(\ref{Mass0}) with the parameters identified as
\be
m\to m_Q,\quad\sigma\to\sigma'=\frac{\sqrt{3}}{2}\sigma_f,\quad \alpha\to C_{\bar{3}}\alpha_S
\ee
to arrive at the mass of the diquark (antidiquark)
\be
m_{(QQ)}=2m_Q+a_0^{(\lambda_Q)}\left(\frac{3\sigma^2}{4m}\right)^{1/3}-C_0^{(QQ)}+\Delta_{SS}^{(QQ)},
\ee
where $\lambda_Q$ is defined in Eq.~(\ref{lambda}) and the hyperfine term (\ref{VSSQ}) gives
\be
\Delta_{SS}^{(QQ)}=\frac{2\pi C_{\bar{3}}\alpha_S}{3m_Q^2}|R_0^{(QQ)}(0)|^2=\frac{2\alpha_S\sigma_f}{9m_Q}\sqrt{\frac{3\beta_Q^3}{\pi}},
\ee
with $\beta_Q$ being a solution of the extremum Eq.~(\ref{difa}) for the diquark (antidiquark) system. 

The constant parameter $C_0$ provides an overall shift of the spectrum. It plays the role of a selfenergy which depends on the quark flavour --- see, for example, the discussion in Ref.~\cite{Kalashnikova:2001ig} and the method to calculate this quantity suggested in 
Ref.~\cite{Simonov:2001iv}. Since the constant mass shift $C_0$ is the same for the same quark and antiquark flavour we set $C_0^{(QQ)}=C_0^{(\bar{Q}\bar{Q})}=C_0^{(Q\bar{Q})}$ and use the values $\Ccc$ and $\Cbb$ found in Ref.~\cite{Kalashnikova:2016bta} --- see Table~\ref{tab:param}.

The values of the auxiliary parameters and the masses of the charmonium and bottomonium diquarks (antidiquarks) obtained as explained above are listed in Table~\ref{tab:QQ}.

\subsubsection{Diquark--antidiquark system}

\begin{table*}[t]
\begin{center}
\begin{tabular}{|c|c|c|c|c|c|c|c|}
\hline
& $\lambda_{(QQ)}$ & $\beta_{(QQ)}$ & $a_0^{\lambda_{(QQ)}}$& $m_{(4Q)}$ (this work), GeV  &  $m_{(4Q)}$, GeV \cite{Karliner:2016zzc} & $m_{(4Q)}$, GeV \cite{Faustov:2020qfm}& $m_{(4Q)}$, GeV \cite{Giron:2020wpx}\\
\hline
$(cc)(\bar{c}\bar{c})$ & 2.95 & 1.75 & -0.92 & 6.196 & 6.1915 & 6.190 & 6.2640-6.2661 \\
\hline
$(bb)(\bar{b}\bar{b})$ & 4.68 & 3.49 & -4.02 & 18.572  & 18.8256 & 19.314& ---\\
\hline
\end{tabular}
\end{center}
\caption{Values of various auxiliary parameters and the masses of the fully-charmed and fully-bottomed $(QQ)(\bar{Q}\bar{Q})$ tetraquarks with $J^{PC}=0^{++}$.}\label{tab:4Q}
\end{table*}

In the developed picture the tetraquark is equivalent to an ordinary quark-antiquark meson with the quark (antiquark) substituted by a antidiquark (diquark) with the only exception that the particles at the ends of the string have spins equal to 1. Then the consideration is similar in spirit to the one for the diquark discussed above, however the parameters of the Hamiltonian (\ref{H}) are now identified as
\be
m\to m_{(QQ)},\quad\sigma\to\sigma_f,\quad \alpha\to C_f\alpha_S,
\ee
where $C_f=4/3$ is the fundamental Casimir operator. In addition, for the tetraquarks with the quantum numbers $0^{++}$ and $2^{++}$, one has 
\bea
&&\langle J_{(4Q)}=0|\veS_{(QQ)}\veS_{(\bar{Q}\bar{Q})}|J_{(4Q)}=0\rangle=-2,\nonumber\\[-2mm]
\\[-2mm]
&&\langle J_{(4Q)}=2|\veS_{(QQ)}\veS_{(\bar{Q}\bar{Q})}|J_{(4Q)}=2\rangle=1.\nonumber
\eea

The predicted masses of the fully-charmed and fully-bottomed scalar tet\-ra\-qu\-arks are listed in Table~\ref{tab:4Q}, where they are also compared with the results of other appro\-ach\-es  which predict the lowest scalar tetraquark near the double-$J/\psi$ threshold. The masses of the tensor ($J^{PC}=2^{++}$) tetraquarks in the spectrum of charmonium and bottomonium calculated in the same model (with all angular momenta between quarks equal to zero) appear to be
\be
m_{(4c)}[2^{++}]=6.56~\mbox{GeV},\quad m_{(4b)}[2^{++}]=18.84~\mbox{GeV}.
\ee 
Thus the tensor fully-charmed tetraquark lies well above the double-$J/\psi$ threshold and therefore is not considered as a candidate for the $X(6200)$ state.

\section{Interplay of quark and molecular dynamics}
\label{sec:interplay}

The findings of the previous chapters can be summarised by stating that the QCD string model predicts the existence of a fully-charmed scalar tetraquark state near the double-$J/\psi$ threshold. The calculated binding energy is
\be
E_B^{(4Q)}=m_{(4c)}[0^{++}]-2m_{J/\psi}\approx -12~\mbox{MeV},
\label{EB}
\ee
where, for consistency, the mass of the $J/\psi$ charmonium was also taken as predicted by the QCD string model (see Table~\ref{tab:cc}). Therefore, in the complex energy plane the compact tetraquark corresponds to a pole just below the double-$J/\psi$ threshold. 

Meanwhile, it is argued in a recent work \cite{Dong:2021lkh} that the interaction between two $J/\psi$ mesons mediated by soft gluons, which hadronise as two-pion and two-kaon exchanges, are capable of forming a near-threshold pole in this system. This implies that a very peculiar situation takes place when two different dynamics independently generate near-threshold poles of the scattering matrix. This case is discussed in detail in a series of papers \cite{Baru:2010ww,Hanhart:2011jz,Guo:2016bjq} and, in particular, it is demonstrated that line shapes of quite a nontrivial form may arise as a result of the interplay of different dynamics present in the system. Below we discuss some consequences of such an interplay in the $X(6200)$. 

Since the energy region covered by this analysis constitutes only several dozen MeV around the double-$J/\psi$ threshold, the coupled-channel problem can be formulated in a simple form,
\be
|X(6200)\rangle=\left(\sqrt{Z}|cc\bar{c}\bar{c}\rangle \atop \chi(\vep)|J/\psi J/\psi\rangle\right),
\label{wf}
\ee
where $|cc\bar{c}\bar{c}\rangle$ and $|J/\psi J/\psi\rangle$ are the compact tetraquark and double-$J/\psi$ molecular state, respectively. The qu\-antity $Z$ introduced by Weinberg \cite{Weinberg:1962hj,Weinberg:1963zza,Weinberg:1965zz} can be interpreted as the probability to observe the state $X(6200)$ in the form of a compact object. The function $\chi(\vep)$ describes the relative motion in the double-$J/\psi$ system. Although originally the Weinberg approach was formulated for a bound state of nucleons --- the deuteron, it was later generalised to other near-threshold states and states from the continuum spectrum as well as to unstable particles --- see Refs.~\cite{Bogdanova:1991zz,Baru:2003qq,Matuschek:2020gqe}. 

The wave function (\ref{wf}) obeys a Schr{\"o}dinger-like equation  (the energy $E$ is counted from the double-$J/\psi$ threshold)
\be
{\cal H}|X(6200)\rangle=E|X(6200)\rangle,
\label{Sheq}
\ee
with the Hamiltonian
\be
{\cal H}=
\left(
\begin{array}{cc}
E_0 &V_{qh}\\
V_{hq}&H_h
\end{array}
\right),
\label{HHH}
\ee
where $E_0$ is the bare energy of the compact state and
\be
H_h(\vep,\vep')=\frac{p^2}{m_{J/\psi}}\delta(\vep-\vep')+V(\vep,\vep').
\label{Vpot}
\ee

The off-diagonal potential $V_{qh}$ describes the transition between the compact tetraquark and the double-$J/\psi$ channel which proceeds via the two string junctions ``annihilation'' followed by the appropriate rearrangement of the released two quarks and two antiquarks to provide two colourless $(\bar{c}c)$ combinations with the proper quantum numbers. This makes difference with the ``'standard'' strong decay mechanism which is provided by a string breaking act resulting in a light quark-antiquark $q\bar{q}$ pair creation from the vacuum --- such a mechanism is clearly responsible for the decays of the compact tetraquark to a doubly-charmed baryon-antibaryon pair or a lighter tet\-ra\-quark plus a heavy-light meson,
\beas
(cc)(\bar{c}\bar{c})&\to& (ccq)+(\bar{c}\bar{c}\bar{q}),\\
(cc)(\bar{c}\bar{c})&\to& (cc)(\bar{c}\bar{q})+(\bar{c}q),\\
(cc)(\bar{c}\bar{c})&\to& (cq)(\bar{c}\bar{c})+(\bar{q}c).
\eeas
The corresponding thresholds lie far away from the energy range under study and therefore are not included into the coupled-channel scheme adopted here.

Near the double-$J/\psi$ threshold the transition form factor can be approximated by a constant,
\be
f_0=\langle cc\bar{c}\bar{c}|V_{qh}|J/\psi J/\psi \rangle.
\label{f0}
\ee

According to the finding of Ref.~\cite{Dong:2021lkh}, potential $V(\vep,\vep')$ supports near-threshold bound or virtual states, so that the corresponding solution of the Lippmann--Schwinger equation,
\be
t_V(\vep,\vep',E)=V(\vep,\vep')-\int \frac{d^3q}{(2\pi)^3}
\frac{V(\vep,\veq)t_V(\veq,\vep',E)}{q^2/(2\mu)-E-i0},
\ee
near the pole can be written in the form 
\be
t_V(\veq,\vep,E)\propto \frac{1}{\gamma_V+ik},
\label{tV}
\ee
where $k=\sqrt{m_{J/\psi}E}$ and $\gamma_V$ is the inversed scattering length, with $\gamma_V>0$ for a bound state and $\gamma_V<0$ for a virtual state.

\begin{figure}[t]
\epsfig{file=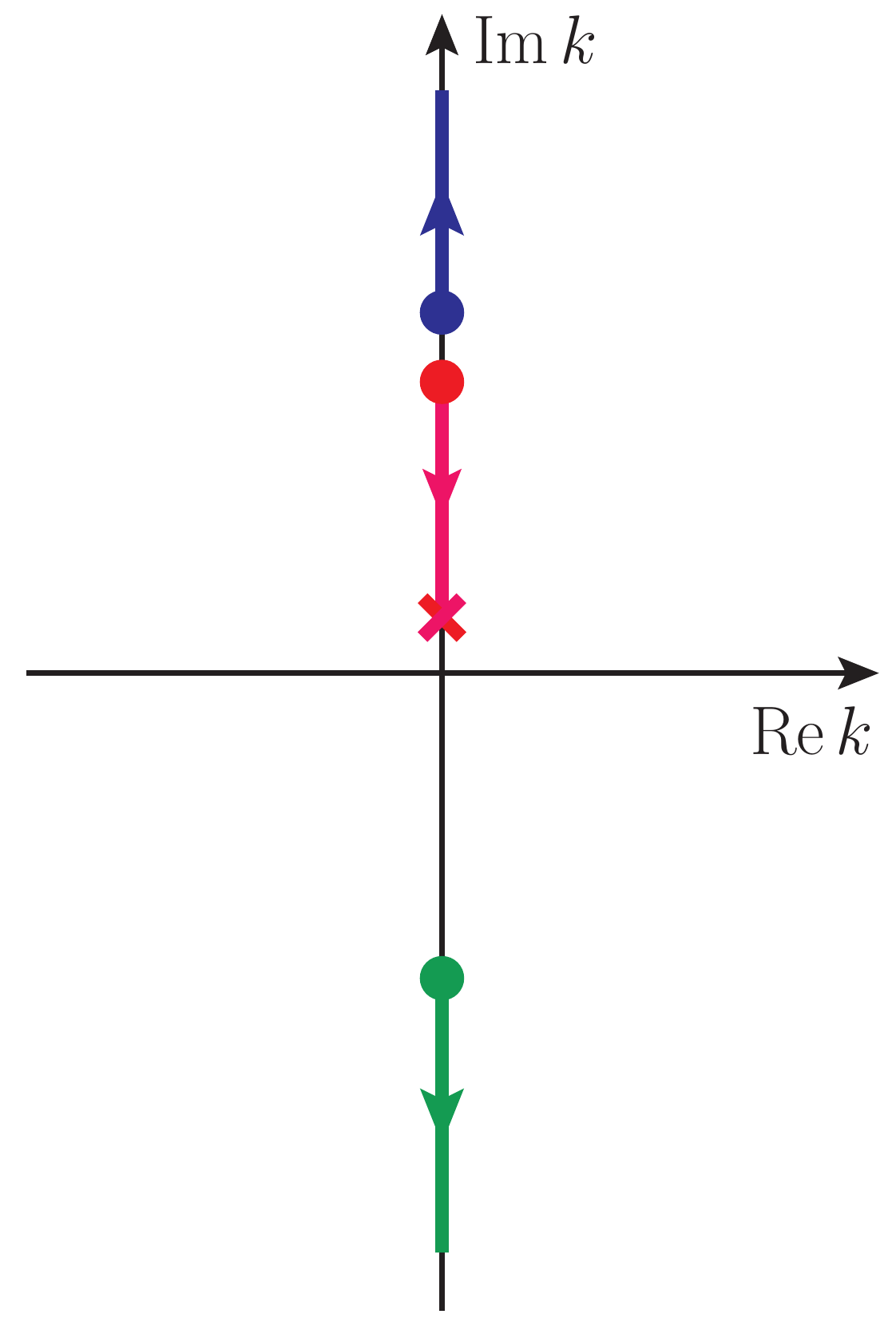, width=0.45\columnwidth}
\epsfig{file=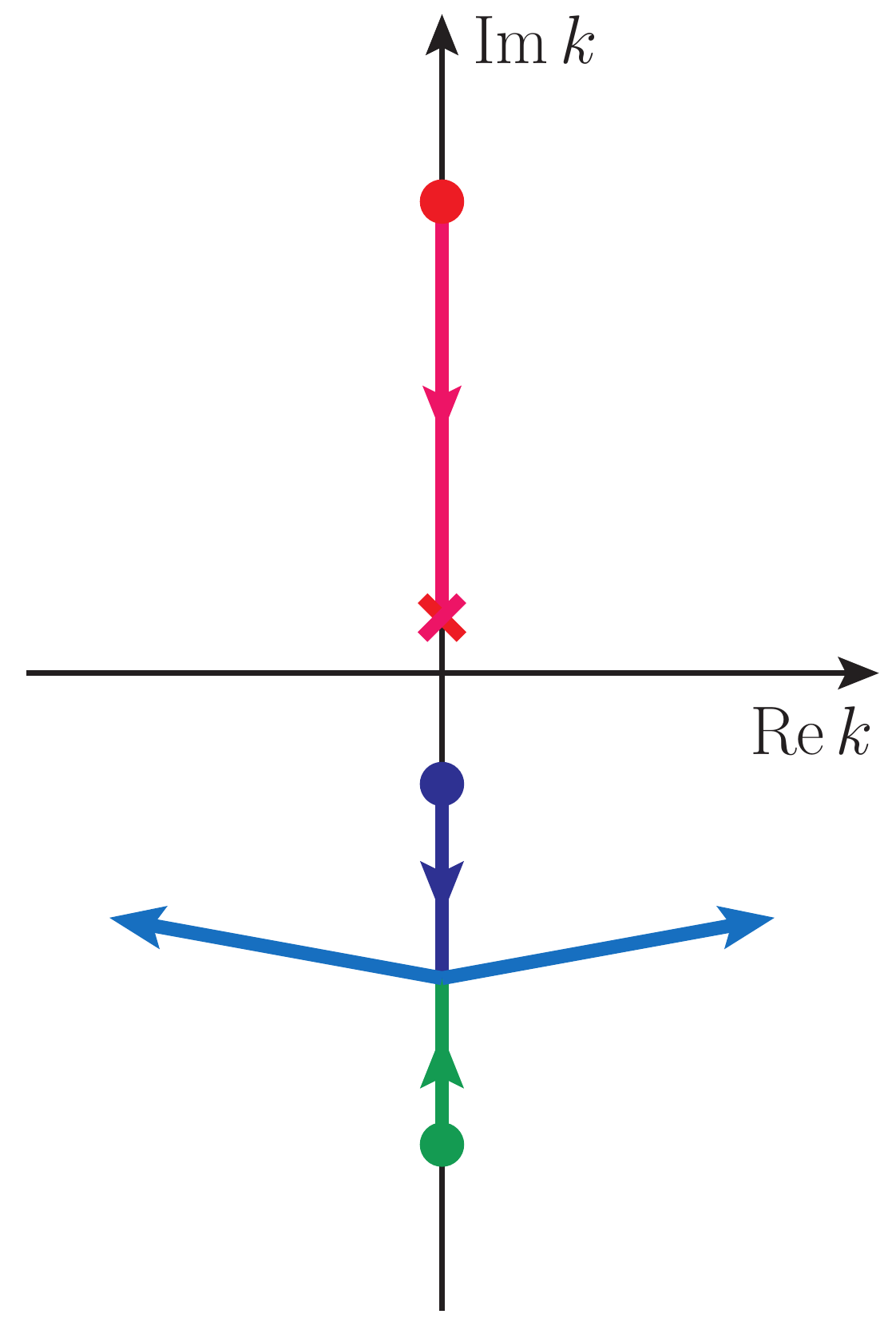, width=0.45\columnwidth}
\caption{Poles motion as the coupling $g_f$ increases starting from zero: the filled circles represent the positions of the bare poles (for $g_f=0$), the arrows show the direction of motion as $g_f$ increases, and the cross represents the physical pole of the $X(6200)$. The left plot corresponds to the case $\gamma_V>0$ (bound state generated by the potential $V(\vep,\vep')$ from Eq.~(\ref{Vpot})) and the right plot is for $\gamma_V<0$ (virtual state generated by $V(\vep,\vep')$).}
\label{fig:poles}
\end{figure}

It is demonstrated in Ref.~\cite{Baru:2010ww} that the scattering matrix in the meson-meson system can be found in the form
\be
t(E)\propto \frac{E-E_f+\frac12 g_f\gamma_V}{(E-E_f)(\gamma_V+ik)+\frac{i}{2}g_f\gamma_Vk},
\label{tmat}
\ee
where $g_f$ is an effective coupling introduced instead of the constant $f_0$ from Eq.~(\ref{f0}) and the parameter $E_f$ defines the bare position of the compact quark state poles, so that we set $E_f=E_B^{(4Q)}<0$ as per Eq.~(\ref{EB}). 

One can easily verify that in the decoupling limit of $g_f=0$ the scattering $t$-matrix (\ref{tmat}) turns to $t^V$ from Eq.~(\ref{tV}). On the other hand, for $g_f\neq 0$, the poles of the scattering matrix in the complex momentum plane can be found as solutions of a cubic equation,
\be
(k^2-m_{J/\psi} E_f)(\gamma_V+ik)+\frac{i}{2}m_{J/\psi} g_f\gamma_V k=0,
\label{eqk}
\ee
so that there are always three poles. In the limit $g_f\to 0$, two of them,
\be
k_{1,2}^{(0)}=\pm i\sqrt{m_{J/\psi} |E_f|},
\label{k12}
\ee
represent the compact tetraquark state and the third one, 
\be
k_3^{(0)}=i\gamma_V,
\label{k3}
\ee 
corresponds to the molecular state. Depending on the relation between $|\gamma_V|$ and $\sqrt{m_{J/\psi}|E_f|}$, either the compact or molecular pole is located closer to the threshold. In Ref.~\cite{Baru:2010ww} these cases are considered separately, however here we do not distinguish between them since we are only interested in the ultimate position of the physical pole responsible for the $X(6200)$ state, no matter from which particular bare pole it originates from.  

As found in Ref.~\cite{Dong:2021lkh}, the two-pion and two-kaon exchanges in the double-$J/\psi$ system are likely to be able to produce a near-threshold pole which represents either a bound or virtual state. We therefore consider both possibilities --- see the left and right plots in  Fig.~\ref{fig:poles}, respectively, where the initial positions of all three bare poles are shown as filled coloured circles and the red cross corresponds to the physical pole (it represents the physical $X(6200)$ as a bound state; the case of a virtual state is discussed below). In the left (right) plot the bare pole located closer to the threshold (irrespective of its nature: whether it is due to the quark or molecular dynamics) is always shown as the red (blue) filled circle while the more remote pole is given by the blue (green) filled circle. The isolated pole (the green and red circles in the left and right plot, respectively) always originates from the compact tetraquark state. As the coupling $g_f$ increases, the poles start to move in the directions indicated by arrows in Fig.~\ref{fig:poles}. Since if two poles collide they are forced to enter the complex plane while this is only allowed in the lower half plane, the cases of $\gamma_V>0$ (left plot) and $\gamma_V<0$ (right plot) demonstrate quite different patterns. It is easy to see that the two plots from Fig.~\ref{fig:poles} exhaust all possibilities compatible with the single physical pole scenario described in Ref.~\cite{Dong:2020nwy}. Namely, in both plots, for a sufficiently large $g_f$ the pole which moves towards the threshold reaches the position of the physical $X(6200)$ pole while the remaining two poles appear to be sufficiently remote. This would not be possible if the bare poles appeared closer to the threshold than the physical pole. Indeed, in this case, for $\gamma_V>0$ (left plot) the physical pole would be reached by the blue circle, while the red one would move close to the threshold to represent a second observable pole, that is a two-pole situation would take place. On the contrary, for  $\gamma_V<0$ (right plot) the physical pole would not be reached at all.  

In the virtual state scenario for the $X(6200)$, its physical pole appears in the lower half plane, and it is easy to convince oneself that again either this pole position cannot be reached by any of the three poles present in the system or one arrives at a multi-pole situation at odds with the findings of Ref.~\cite{Dong:2020nwy}. 

In short, in the single-pole scenario for the $X(6200)$ its physical pole is necessarily hit by the closest to the threshold pole moving towards it, as depicted in Fig.~\ref{fig:poles}. Furthermore, it is possible to discriminate between these two scenaria. To this end we check the physical contents of the state described by the physical pole. To reduce the number of parameters, without loss of generality, we study two cases for $\gamma_V$, 
\be
\gamma_V=\pm \sqrt{m_{J/\psi} |E_f|},
\label{gVE}
\ee
that implies that in the zero-coupling limit the pole due to the direct interaction in the double-$J/\psi$ system coincides with one of the compact tetraquark poles. Both assumptions from Eq.~(\ref{gVE}) comply well with the results reported in Ref.~\cite{Dong:2021lkh}. Then the only remaining free parameter --- the coupling $g_f$ --- defines not only the current position of the three poles but also the value of the Weinberg $Z$-factor (details of its derivation can be found in Ref.~\cite{Baru:2010ww}). As a representative example we consider 
\be
k_{X(6200)}^{\rm phys}=i50~\mbox{MeV}
\ee
that sets up the position of the physical pole in the momentum complex plane (the red cross in Fig.~\ref{fig:poles}). This corresponds to the mass
\be
M_{X(6200)}^{\rm phys}\approx 6193~\mbox{MeV}
\ee
consistent with the findings of Ref.~\cite{Dong:2020nwy}.

The results of calculations are collected in Table~\ref{tab:results}. One can easily see from these results that the two options for the molecular dynamics in the double-$J/\psi$ system result in utterly different physical states. Indeed, although the values of the coupling $g_f$ needed to reach the physical pole are the same for both of them, the Weinberg's $Z$-factor is different: $Z\ll 1$ for $\gamma_V>0$ but $Z\sim 1$ for $\gamma_V<0$. This implies that only the bound-state scenario with $\gamma_V>0$ is compatible with the results of work \cite{Dong:2020nwy} where the compositeness of the $X(6200)$ is found to be close to unity. In other words, it is argued there that the data currently available prefer the $X(6200)$ as a compound (molecular) rather than compact object, the compact component of the wave function being small. Since the physical interpretation of the $Z$-factor is the probability to observe the studied resonance as a compact state, the data clearly prefer the case $Z\ll 1$, that is $\gamma_V>0$. An additional argument against the virtual state scenario ($\gamma_V<0$) is provided by the position of the zero of the scattering amplitude in the double-$J/\psi$ system --- see the last column of Table~\ref{tab:results}. This quantity was introduced in Ref.~\cite{Baru:2010ww} as
\be
E_C=E_f-\frac12 g_f\gamma_V.
\label{EC}
\ee 

From Eq.~(\ref{tmat}) one can see that the scattering amplitude of the two $J/\psi$'s off each other vanishes at $E=E_C$. Since the wave function of the $X(6200)$ has two components, it can be produced through both of them, with the full production amplitude being a weighted mixture of the two amplitudes which correspond to the production via the compact and molecular components. Meanwhile, it was found in Ref.~\cite{Dong:2020nwy} that the LHCb data can be well described if the production source is associated with the double-parton scattering (DPS) process that implies the production through two jets which further hadronise into two $J/\psi$ charmonia. In other words, the production through the molecular component of the $X(6200)$ is preferred, the latter vanishing at $E=E_C$, if $E_C$ appears in the near-threshold region above threshold. Since the single-parton scattering (SPS) amplitude, that is production though the compact tet\-ra\-qu\-ark component of the wave function (which does not have zero at $E_C$ --- see Ref.~\cite{Baru:2010ww}) may also somewhat contribute to the total production amplitude the line shape may have a sort of dip at $E=E_C$ rather than a strict zero, however such an irregular behaviour would have a footprint in the data. As one can see from Table~\ref{tab:results}, $E_C<0$ for the bound state scenario, so that no additional structures in the data above threshold are expected to occur. On the contrary, in the virtual state scenario, $E_C>0$ that appears to be at odds with the data which do not demonstrate any kind of structures just above the double-$J/\psi$ threshold. 

We conclude, therefore, that the finding of Ref.~\cite{Dong:2020nwy} together with the existence of a compact tetraquark state just below the double-$J/\psi$ threshold, as found in this work, limit the set of options available for the molecular dynamics in the double-$J/\psi$ system as well as for the nature of the $X(6200)$. Namely, only the bound state scenario for both the bare pole in the  double-$J/\psi$ system and the physical $X(6200)$ pole is favoured by the data currently available.

\begin{table}[t]
\begin{center}
\begin{tabular}{|c|c|c|c|c|c|c|c|}
\hline
Scenario & $g_f^{(0)}$ & $Z$ & $E_C$, MeV \\ 
\hline
$\gamma_V>0$ & 0.33 & 0.1 & -44 \\
$\gamma_V<0$ & 0.33 & 0.4 & 20\\
\hline
\end{tabular}
\end{center}
\caption{The value of the coupling $g_f$ when the physical pole location is reached by one of the three poles of the scattering matrix and the corresponding values taken by the Weinberg $Z$-factor and the zero of the scattering amplitude $E_C$ for both scenaria ($\gamma_V>0$ and $\gamma_V<0$) for the molecular dynamics in the double-$J/\psi$ system.}\label{tab:results}
\end{table}

\section{Discussion}
\label{sec:discussion}

In this work we evaluated the mass of the lowest fully-charmed tetraquark state in the framework of the QCD string approach. The calculation done is parameter-free and can be regarded as a pure prediction of the model. The values of the parameters used were previously totally fixed to provide the best overall description of the spectrum of the low-lying generic $\bar{c}c$ charmonia. The result obtained appears in a good agreement with similar predictions found in the literature and provides a strong candidate for the (possibly present) compact component of the state $X(6200)$ predicted recently from the theoretical coupled-channel analysis of the LHCb data on the double-$J/\psi$ production in proton-proton collisions. The quantum number $0^{++}$ are preferred since the tensor fully-charm\-ed tetraquark candidate lies well above the double-$J/\psi$ threshold. Predictions for other, excited, fully-charmed tetraquarks can be made in the same framework as soon as experimental data require them. 

A prediction for the mass of the lowest fully-bottomed tetraquark is also provided, however it lies about 300 MeV below the double-$\Upsilon(1S)$ threshold and for this reason can hardly play a role for the formation of a near-threshold pole, if it exists in analogy with the double-$J/\psi$ system. This difference between the charmonium and bottomonium tetraquarks should not come as a surprise given the different dynamical regimes realised in them, as contained in Eq.~(\ref{mcm0mb}). In principle, an excited fully-bottomed tetraquark could potentially have a bare mass located near one of the double-$\Upsilon$ thresholds, however this investigation lies beyond the scope of this work.

The currently available data on the double-$J/\psi$ production in proton-proton collisions are consistent with the existence of a pole near the double-$J/\psi$ threshold whose nature cannot be completely understood at the moment. If the wave function of the corresponding state $X(6200)$ contains a compact component, a fully-charmed tetraquark discussed in this work is a good candidate for this role. Meanwhile, since the direct interaction in the double-$J/\psi$ system is also very likely to generate a near-threshold pole then a nontrivial interplay of the quark and meson degrees of freedom takes place in the $X(6200)$ which was studied in this work using the formalism developed in Ref.~\cite{Baru:2010ww}. The results obtained, consistent with the scenario of a single near-threshold pole concluded in Ref.~\cite{Dong:2020nwy} from the coupled-channel analysis of the LHCb data, favour the physical $X(6200)$ to be a shallow bound state. Further steps in identifying the nature of this state may be possible after additional data, including measurements in complementary channels, as mentioned, for example, in Ref.~\cite{Dong:2020nwy}, become available.

\section*{Acknowledgments}
This work was supported by the Ministry of Science and Education of Russian Federation (grant 14.W03.31.0026).

\end{document}